\documentclass[12pt,reqno]{article}

\def\fnote#1#2{\begingroup\def\thefootnote{#1}\footnote{#2}\addtocounter{footnote}{-1}\endgroup}

\usepackage{hyperref}
\hypersetup{
    colorlinks=true, linkcolor=blue, filecolor=red, urlcolor=cyan, 
    pdftitle={Overleaf Example}
    pdfpagemode=FullScreen,
    }
\usepackage{url}

\usepackage{amsmath}
\usepackage{makeidx}
\usepackage{epsf}
\usepackage{graphicx}        
\usepackage{caption, float}
\usepackage{cite}

\makeindex

\def\inbar{\vrule height1.5ex width.4pt depth0pt}
\def\IB{\relax{\rm I\kern-.18em B}}
\def\IC{\relax\,\hbox{$\inbar\kern-.3em{\rm C}$}}
\def\ID{\relax{\rm I\kern-.18em D}}
\def\IE{\relax{\rm I\kern-.18em E}}
\def\IF{\relax{\rm I\kern-.18em F}}
\def\IG{\relax\,\hbox{$\inbar\kern-.3em{\rm G}$}}
\def\IH{\relax{\rm I\kern-.18em H}}
\def\II{\relax{\rm I\kern-.18em I}}
\def\IK{\relax{\rm I\kern-.18em K}}
\def\IL{\relax{\rm I\kern-.18em L}}
\def\IM{\relax{\rm I\kern-.18em M}}
\def\IN{\relax{\rm I\kern-.18em N}}
\def\IO{\relax\,\hbox{$\inbar\kern-.3em{\rm O}$}}
\def\IP{\relax{\rm I\kern-.18em P}}
\def\IQ{\relax\,\hbox{$\inbar\kern-.3em{\rm Q}$}}
\def\IR{\relax{\rm I\kern-.18em R}}
\def\IT{\relax{\rm I\kern-.18em T}}
\def\ZZ{\relax{\sf Z\kern-.4em Z}}

\def\nablaslash{\relax{\rm /\kern-.28em \nabla}}

\def\a{\alpha}   \def\b{\beta}    \def\g{\gamma}  
\def\e{\epsilon}    \def\k{\kappa}  
\def\L{\Lambda}     \def\si{\sigma}

\def\cA{{\cal A}}

\def\cO{{\cal O}} \def\cP{{\cal P}}  \def\cR{{\cal R}}

 \def\oD3{{\overline \rmD 3}}

\def\adot{{\dot{a}}}

\def\fnote#1#2{\begingroup\def\thefootnote{#1}\footnote{#2}\addtocounter
{footnote}{-1}\endgroup}
\def\beq{\begin{equation}}
\def\eeq{\end{equation}}
\def\bea{\begin{eqnarray}}
\def\eea{\end{eqnarray}}
\def\llea#1{\label{#1}\eea}
\def\lleq#1{\label{#1}\eeq}
\let\nn=\nonumber
\def\tabroom{\hbox to0pt{\phantom{\Huge A}\hss}}
\def\notin{\ \hbox{{$\in$}\kern-.51em\hbox{/}}}

  \def\E1Fq{E_1/\IF_q}

\def\rmD{{\rm D}}

\def\rmSL{{\rm SL}}

 \def\rmD{{\rm D}}

     \newcommand{\rmpiv}{{\rm piv}}

     \def\rmrh{{\rm rh}}

          \def\rmGeV{{\rm GeV}}       
\newcommand{\rmHI}{{\rm HI}}      
\newcommand{\rmHSI}{{\rm HSI}}       
\newcommand{\rmIII}{{\rm III}}

        \def\rmPl{{\rm Pl}}

\def\rmSL{{\rm SL}}

\def\notdiv{{\relax{~|\kern-.35em /~}}}

\def\boxit#1{
\vbox{\hrule height1pt\hbox{\vrule width1pt\kern0.3cm
\vbox{\kern0.3cm\hbox{$\displaystyle#1$}\kern0.3cm}\kern0.3cm\vrule
width1pt}\hrule height1pt}}

\begin{document}
\parindent=0pt

\phantom{\hfill {\bf Draft} \break
\phantom{whatever} \hfill \today~}

\vskip .8truein

\centerline{\large {\bf Scaling Characteristics of Hilltop and Hilltop-Squared Inflation}}

\vskip .4truein

\centerline{\sc Monika Lynker\fnote{1}{mlynker@iu.edu} and 
   Rolf Schimmrigk\fnote{2}{rschimmr@iu.edu, netahu@yahoo.com}}

\vskip .3truein

\centerline{Dept. of Physics}

\vskip .1truein

\centerline{Indiana University South Bend}

\vskip.1truein

\centerline{1700 Mishawaka Ave., South Bend, IN 46615}

\vskip 1.3truein

\baselineskip=17pt

\centerline{\bf Abstract}

One of the longstanding goals in the framework of inflation is the construction of tools that can be used to 
classify models in theory space. An idea that has been put forward in this context is to consider the energy dependent 
scaling behavior of observables to characterize different models.
We implement this approach in the framework of hilltop and hilltop-squared inflation by analyzing their observables when the 
small-field approximation is not imposed and the energy scale $\mu$ of these models is varied as a free 
parameter, subject to observational constraints. 
 We show that the scalar spectral tilt  and the tensor ratio $r$ exhibit $\mu$-dependent scaling behavior 
 and that the scaling exponents  as functions of $\mu$ in turn lead to functional forms that are model dependent.
 Scaling relations of the type discussed here are of interest as characteristics of the inflationary theory space 
 as well as in the context of the post-inflationary reheating process.
  We further observe a bifurcation behavior in the behavior of $p$-families in the spectral-tensor plane for 
a critical value of $\mu$.

\renewcommand\thepage{}
\newpage
\parindent=0pt

\pagenumbering{arabic}

\baselineskip=18pt
\parskip=0.02truein

\tableofcontents

\vfill \eject

\baselineskip=19.5pt
\parskip .1truein
\parindent=0pt

\section{Introduction}

Hilltop type potentials have remained among the canonical models of  inflation. 
While the general class of  hilltop theories encompasses a wide variety of potentials that have 
been discussed in the context of a number of embedding theories, see e.g. the reviews \cite{lr98, mrv13} 
for many references, our focus here is  on a more specific set of models, defined as 
\beq
 V_{p,n} ~=~ \L^4 \left( 1 ~-~ \left(\frac{\phi}{\mu}\right)^p\right)^n.
 \lleq{hi+hsi}
 where $n=1,2$.
 Some of the models in this class  have become benchmark models for the CMB collaborations 
    \cite{a15etal, cmbs4first, a18etal, simons18, a19etal, a21etal}.  While hilltop models in general, and the $V_{p,n}$ above 
  in particular, have traditionally been considered in the context of singlefield inflation (see e.g.  
     \cite{mrv13, km95a, km95b, l96, bl05, kr05, mr06, kll07, b08etal, mr10, a10etal, ep11, r13, gr14, bpk16, dpb16,  l18, 
            c19etal, kl19, gy19,  ads19etal, d20, g20, o21etal, hs21, c22, lr22, g23etal}),
 this class of potentials is also of interest in the framework of multifield inflation models that describe saddle point inflation.  
In this context hilltop potentials can be used  as singlefield approximations of inflaton 
trajectories that roughly evolve along a cross-section in the concave direction of the multifield potentials. 
Examples of this latter type include the models of $j$-inflation and $h$-inflation
 in the framework of modular inflation \cite{rs14, rs15, rs16, rs17, ls19, rs21}.
 
 The potentials $V_{p,n}$ describe two-parameter singlefield inflationary models and as such are amenable to the scaling 
 type analysis that was  introduced in ref. \cite{ls22} in the context of multifield inflation. In was observed there that theories with 
 two parameters  may lend themselves to a characterization in terms of a family of scaling relations with associated 
 scaling exponents $\b_p$ that are specific to the models and therefore could be used to distinguish regions 
 in the inflationary theory space. Furthermore, if the exponents $\b_p$ are functions $\b_p(\mu)$
 of the energy scale $\mu$ one can ask whether they in turn show scaling behavior with an associated exponent $\g_p$
 that would then provide a single number associated to the parameter space of the model.
  Such scaling relations were established in \cite{ls22}  in the context 
 of modular inflation, which is a class of twofield models. The same question can be raised in the context of other
 models with two parameters and in the present analysis we focus on hilltop type potentials. We vary 
 the exponent $p$ in $V_{p,n}$ systematically and consider the dependence of the observables on the energy scale $\mu$.
 We will focus on pure slow-roll inflation, without making any further approximations that were adopted 
 in early papers on these models.  Making such additional approximations eliminates the dependence of the observables on the 
 energy  and thereby  trivializes the scaling behavior.  Imposing such approximations implies, for example, that 
 the critical exponent of the tensor ratio is energy independent, hence there is no functional dependence
  of the scaling exponents on $\mu$.
 
Depending on the exponent $n$ in (\ref{hi+hsi}) the models exhibit a quite different structure close to the minimum at $\phi=\mu$. 
While hilltop inflation ($n=1$)  intersects the zero line with a non-zero slope, hilltop-squared models have a smooth valley around the minimum.
The fact that there is no valley for the hilltop models has sometimes been considered a weakness of these models because strictly speaking this means 
that these model do not have a domain that allows the inflaton 
 to oscillate around the minimum, thereby releasing its energy to other particles whose couplings to the inflaton 
 are assumed to be small and negligible during inflation but become relevant during the post-inflation process
  of reheating.
 As a practical matter this issue can be resolved by simply gluing an appropriate potential to 
$V(\phi_e)$ for $\phi> \phi_e$ to the hilltop potential.
 The fact that this issue does not arise for hilltop-squared theories has been a motivation to consider 
 these models, see e.g. \cite{a19etal, hs21, lr22, g23etal}.

We have organized this paper as follows. 
Section 2 describes the observables we will focus on in both the pure slow-roll
     and in their different small-field approximations. 
In section 3 we briefly consider the reheating constraint so as to be able to gauge its 
    effect when imposed in addition to the CMB constraints. 
In section 4 we present the impact of the CMB and reheating constraints for the class of hilltop models and 
in section 5 we discuss the scaling distributions that will be made quantitative for hilltop models in section 6. 
In section 7 we analyze $p$-families of hilltop models in dependence of the energy scale $\mu$ and describe 
     a bifurcation phenomenon that appears at a critical value of this scale.
In section 8 we extend our analysis to the hilltop-squared models and 
in section 9 we discuss the implications of our scaling analysis as a tool that provides characteristic 
quantities defined on the inflationary landscape.  
We conclude in section 10.
 
\section{Observables in hilltop inflation}

Our focus in this work is on hilltop inflation in the pure slow-roll approximation but it is of interest to compare our 
results with the behavior of this class of models in the small-field approximation. We discuss these approximations 
it turn.

\subsection{Slow-roll approximation}

The analysis in the following is concerned with the observables at horizon crossing as functions of the inflaton
 $\phi_*$ and the number of e-folds $N_*$ between horizon crossing time $t_*$ and the end of inflation $t_e$. 
 For the Lukash-Bardeen perturbation   \cite{l80, b80, w08}
 \beq
 \cR ~=~ H\delta u ~-~ \psi,
 \eeq
 where $H=\adot/a$ is the Hubble-Slipher parameter, $\psi$ is the metric potential, and $\delta u$ is obtained 
 from $\delta T_{0i}$, the spectral index and the tensor-to-scalar ratio take the form
  \bea
 n_{\cR\cR}
    &=& 1 - p\frac{M_\rmPl^2}{\mu^2} \frac{ \left(
      \frac{\phi}{\mu}\right)^{p-2}}{\left(1- \left(\frac{\phi}{\mu}\right)^p\right)^2} \left[2(p-1) + (p+2) \left(\frac{\phi}{\mu}\right)^p\right]\nn \\
  r &=&  8p^2 \frac{M_\rmPl^2}{\mu^2}  \frac{ \left(\frac{\phi}{\mu}\right)^{2(p-1)} }{
   \left(1-\left(\frac{\phi}{\mu}\right)^p\right)^2}.
   \llea{HI-observables}
 while the number of e-folds is given by
 \beq
  N_*  ~=~ \frac{\mu^2}{pM_\rmPl^2} 
    \left[ \frac{1}{p-2}
     \left( \left(\frac{\mu}{\phi_*}\right)^{p-2}  ~-~ \left( \frac{\mu}{\phi_e}\right)^{p-2} \right) 
    - \frac{1}{2} \left( \left(\frac{\phi_e}{\mu}\right)^2 -  \left(\frac{\phi_*}{\mu}\right)^2\right) \right]
  \lleq{sr-efolds}
  The end-of-inflation value $\phi_e$ will be determined by the constraint $\e_V(\phi_e)=1$, which is obtained
  from the constraint
  \beq
   \left(\frac{\phi_e}{\mu}\right)^p ~+~ \frac{p}{\sqrt{2}} \frac{M_\rmPl}{\mu} ~\left(\frac{\phi_e}{\mu}\right)^{p-1} ~-~ 1 ~=~ 0.
  \lleq{phie-via-epsV}
  
We will use two different ways to analyze these models phenomenologically.
The first is the ``scan method", in which the strategy is to systematically scan the parameter space $(\L, \mu, \phi_*)$ for viable 
 models within the available constraints, in particular the CMB  results for $n_{\cR\cR}, r$ from {\sc Planck} as well as
     the adopted range for the number of e-folds. This is in the spirit of earlier analyses in a different context, such as \cite{rs21}.
 The second is the ``$p$-family method", where the idea is to choose a  fixed number of e-folds $N$, say $N=60$, and 
    determine $\phi_*$ in terms 
     of $N_*$, where  $\phi_e$ has been solved via $\e_V(\phi_e)=1$,   as usual.
    Having determined $\phi_*(N)$ by inverting the $N_\star$ equation in (\ref{sr-efolds}) leads to
    \beq
     \left(\frac{\phi_*}{\mu}\right)^p ~-~ A_p  \left(\frac{\phi_*}{\mu}\right)^{p-2} ~+~ \frac{2}{p-2} ~=~ 0
    \lleq{phistar-via-N}
    with 
    \beq
     A_p ~:=~ 2p \frac{M_\rmPl^2}{\mu^2} N_* ~+~  \left(\frac{\phi_e}{\mu}\right)^2 ~+~ 
      \frac{2}{p-2} \left(\frac{\mu}{\phi_e}\right)^{p-2}.
     \eeq
   In this way  one can compute the observables  for a given $\mu$.
    This method was considered for example in the early hilltop papers \cite{km95a, km95b} (see also \cite{gr14}).

\subsection{Small-field approximations of hilltop inflation}

The small-field approximation can be implemented in different forms. 
The weakest form requires only that the 
initial value $\phi_*$ is much smaller than the energy scale $\mu$. More interesting is the stronger approximation that 
$\phi$ is much smaller than $\mu$ throughout inflation, because this allows to obtain analytical expressions for the observables. 
 While consistency with the model does place restrictions  on $\mu$, these are
model dependent and are weaker than the sub-Planckian constraint often encountered as part of the small-field definition.
 In order to distinguish these different 
approximations we use the designations SF.I for the weakest version, SF.II for the version that covers all of inflation, 
and SF.III for the version that furthermore imposes an a priori constraint on the energy scale $\mu$ that is 
model independent.

The motivation for SF.II comes from the fact that in this case one can use the resulting e-fold formula
 \beq
  N_* ~=~ \frac{\mu^2}{p(p-2)M_\rmPl^2} 
     \left( \left(\frac{\mu}{\phi_*}\right)^{p-2}  ~-~ \left( \frac{\mu}{\phi_e}\right)^{p-2} \right) 
\eeq
 to solve analytically for $\phi_*$ as a function of $N_*$. 
Using this solution we can write the e-fold dependence of the spectral index as
\beq
 n_{\cR\cR}(p,\mu,N_*) ~=~  1 - 
   \frac{2(p-1)  }{(p-2) N_* ~+~\frac{1}{p} \left(\frac{p}{\sqrt{2}}\right)^{(p-2)/(p-1)}  \left(\frac{\mu}{M_\rmPl}\right)^{p/(p-1)}}.
 \lleq{specind-sfII}
 and the tensor-to-scalar ratio as
 \beq
 r(N_*,\mu,p) 
  ~=~ \frac{8p^2 \left( \frac{\mu}{M_\rmPl}\right)^{2p/(p-2)}   }{\left( p(p-2) N_*  
    + \left(\frac{\mu}{M_\rmPl}\right)^{p/(p-1) } 
           \left(\frac{p}{\sqrt{2}} \right)^{(p-2)/(p-1)}
            \right)^{2(p-1)/(p-2)} },
  \eeq
This $\mu$-dependent form of the small-field approximation does not lead to an analytical $N_*$-scaling relation for either of the observables, 
which is presumably why it has not been considered in the literature. However, we can think of these expressions as 
functions that indicate a kind of ``effective" scaling relation with energy dependent exponents and amplitudes. This 
point will turn out to be useful later in this paper to gain a qualitative understanding of the $\mu$-dependence 
of hilltop inflation.

While neither $r$ nor $n_{\cR\cR}$ scale with $N_*$ directly in the SF.II approximation, 
the tensor ratio does scale analytically with the spectral tilt  $\delta_n=1-n_{\cR\cR}$ as 
  \beq
   r(n_{\cR\cR},\mu) ~=~ \frac{8p^2\left(\frac{\mu}{M_\rmPl}\right)^{2p/(p-2)}}{(2p(p-1))^{2(p-1)/(p-2)}}~
  \delta_n^{2(p-1)/(p-2)}.
  \lleq{SFII-r-vs-delta}
This shows that even in the SF.II approximation the tensor ratio varies for the class of hilltop models and that typicality 
statements made in the literature  about the behavior of the tensor ratio as a function of $n_{\cR\cR}$ do not apply in general.
The slow-roll form of this relation will be discussed in section 4.

 Our results below will show that while the SF.II approximation does capture qualitatively certain 
  features of hilltop inflation it is only a good approximation in a limited parameter range that is not enforced by observations.
Furthermore, while in some parameter regions the small-field approximation might be reasonable at the beginning
of hilltop inflation it is in general less so toward the end of inflation.  
If the $\mu$-terms in the denominators are neglected the spectral  index reduces in the resulting SF.III approximation to 
the $\mu$-independent spectral index 
 \beq
 n_{\cR\cR}(p,N_*) ~=~ 1 - \frac{2(p-1)}{(p-2)} \frac{1}{N_*}
 \lleq{specind-triple-approx}
  considered in the early hilltop papers \cite{km95a, km95b}. Similarly, in this limit the tensor ratio 
 reduces to the form discussed in  \cite{l96} 
  \beq
   r(N_*) ~=~ 8p^2 \left(\frac{\mu}{M_\rmPl}\right)^{2p/(p-2)} \frac{1}{\left(p(p-2) N_*\right)^{2(p-1)/(p-2)}}.
   \lleq{tensor-ratio-triple-approx}
   In both relations the exponents are independent of the energy, which is not the case in the slow-roll or 
   the small field approximation SF.II.
   
\vskip .3truein

\section{Reheating constraint}

While our main focus is in the energy dependence of the scaling behavior in hilltop and hilltop-squared inflation, it is 
of interest to see how the reheating constraints affect the distributions associated with the initial conditions of the inflaton 
field. For this we consider the usual insertion of phases between horizon crossing and today, in combination with the  evolution 
of the energy density during reheating and the conservation of entropy between the end of reheating and today. The simplest
analyses assume that the equation of state parameter $w_\rmrh$ is constant, so that
\beq
\rho_\rmrh ~=~ \rho_e \left(\frac{a_e}{a}\right)^{3(1+w_\rmrh)},
\eeq
  an assumption that is made in much of the literature.
  The detailed form of the phase evolution analysis of the scale factor depends on how the expansion of the $k_*/k_0$ is 
  done but the end-result  is  essentially the same, see for example \cite{dkw14, mk14, glp15, cgw15, c15etal}. 
  For our purpose it is useful to evaluate the resulting formula for the 
e-folds $N_\rmrh$ during reheating for potentials of the type $V=\L^4 f(\phi/\mu)$, where $f$ is a dimensionless function. The 
overall energy scale $\L$ is determined by the CMB amplitude $\cA_\cR$, leading to 
\beq
 N_\rmrh ~=~ \frac{4}{1-3w_\rmrh} \left(-N_*  ~+~ \frac{1}{4} \ln \frac{(f'_*)^2}{f_*f_e} ~+~ \frac{1}{2} \ln \frac{M_\rmPl}{\mu} 
 ~+~ \si_\rmpiv \right),
 \lleq{Nrh-for-general-f}
 where $f_*$ is the dimensionless potential evaluated at $\phi_*$, $f_e$  is its value at the end-of-inflation, and
  $f'$ is the dimensionless derivative of $f$. The parameter  $\si_\rmpiv$ is a model independent constant
 \beq
  \si_\rmpiv ~=~ \ln \frac{a_0T_0}{k_*} ~-~ \frac{1}{3}\ln \frac{11g_\rmrh}{43} - \frac{1}{4} \ln \frac{405}{\pi^2 g_\rmrh} 
    + \frac{1}{4} \ln (12\pi^2 \cA_\cR),
  \eeq
  determined by the CMB amplitude  $\cA_\cR$, the pivot scale $k_*$, as well as $g_\rmrh$ and $T_0$.
  For the reheating temperature we obtain in terms of the dimensionless potential $f$ 
  \beq
   T_\rmrh ~=~ M_\rmPl \left(\frac{540}{g_\rmrh} \frac{M_\rmPl^2}{\mu^2} \cA_\cR 
     \frac{(f'_*)^2 f_e}{f_*^3}\right)^{1/4} ~
        e^{-\frac{3}{4}(1+w_\rmrh)N_\rmrh}.
   \lleq{Trh-for-general-f}
 This compact form is convenient for computations even though it is written in a slightly redundant form.
  The energy dependent scaling relations obtained below for the observables 
  lead to scaling behavior of $N_\rmrh$ and $T_\rmrh$ in 
 dependence of these variables.  
 
\vskip .2truein

\section{Scaling behavior of hilltop inflation}

Hilltop inflation has mostly been analyzed at low $p$ and low $\mu$. 
The model at $p=4$ in particular has been discussed often in the literature, see e.g. 
 \cite{km95a, km95b, l96, bl05, mrv13, l18, c19etal, kl19, d20, g20, hs21, c22}, 
and has been adopted as a bench-mark model by the CMB collaborations 
(see e.g. \cite{a15etal, cmbs4first, a18etal, a19etal, a21etal}). 
In this section we investigate the impact of the CMB and reheating constraints 
more generally for these models and in particular analyze the behavior of the observables as the energy 
parameter $\mu$ is varied, similar to the analysis in \cite{ls22} in the context of multifield inflation. The resulting 
data will  be used in later sections for the scaling analyses. 

We vary $p$  through a wide set of models, ranging from 3 to 100, and for each $p$  consider  a range of $\mu$ values that
 is compatible with the CMB constraints on the spectral index $n_{\cR\cR}$ and 
 the tensor-to-scalar ratio $r$,  and also lead  to sufficient inflation $N_* \in [50,70]$. 
 This $\mu$-range increases as $p$ gets larger, both in the sub-Planckian and the super-Planckian regime. 
 For low $p$ in particular the models are not consistent with the CMB and e-folds constraints  
 for sub-Planckian values for $\mu$, but these become viable as $p$ increases. 
We scan the complete $\mu$-range for super-Planckian values in an integral parametrization 
$\mu=mM_\rmPl$, $m$ an integer, and impose for larger $p$ the generic $p-$independent 
lower bound $\mu=10^{-2} M_\rmPl$ for the sub-Planckian range. This lower bound is sufficient to indicate 
the behavior of these models, 
in particular toward the low end of the tensor ratio that are probed by such $\mu$-values.  
The parametrization of these sub-Planckian ranges is via rational values $\mu=M_\rmPl/m$. The upper bound on 
$\mu$ grows linearly with $p$ with a slope of slightly less than eight, while  the full sub-Planckian regime  reaches for 
large $p$ far below the Planck scale. While there is a weak correlation between the number of inflationary e-folds and 
 $\mu$, there is quite some degeneracy even for low $p$ and this degeneracy increases for larger $p$.
 For the purpose of our discussion a full scan of $\mu$-range is not necessary.

  In Fig. 1 we illustrate the distributions mapped out   by viable initial conditions
   in the spectral-tensor plane considered by the experimental collaborations. 
   We adopt the values $n_{\cR\cR}\in [0.96, 0.97]$ and $r_*\leq 0.035$, as suggested by the CMB experiments \cite{a21etal}.
   The color coding is blue for super-Planckian $\mu$-range and pure inflationary constraints, while red indicates 
   the results after taking into account the  reheating constraint $N_\rmrh$, which is implemented  here for illustration
    with $w_\rmrh=0$.  Black is for  $\mu$ in the sub-Planckian range with inflationary constraints and yellow indicates
    the  inclusion of the reheating constraint.
   The first panel in Fig. 1 shows that the quartic hilltop model has no viable initial values for sub-Planckian $\mu$-values, 
   which also holds for $p=3$. The ranges of the inflaton field in both models are such that these models are 
   no longer small-field inflation, given the current observational constraints.
   As $p$ increases sub-Planckian energy values become possible and the viable $\mu$-bounds extend both in the 
   sub- and the super-Planckian regime, leading to the regions in the parameter space in which these models can 
   describe 
   small-field inflation.
 
 \begin{center}
 \includegraphics[scale=0.7]{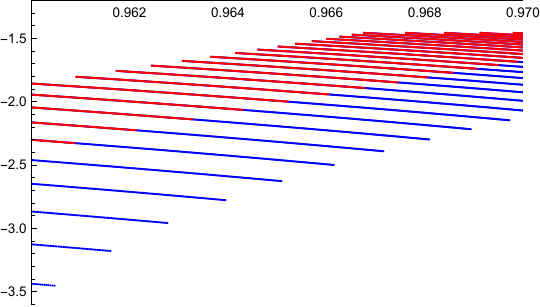}
~
 \includegraphics[scale=0.7]{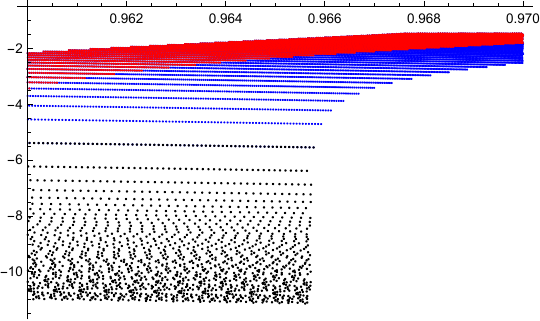}
~
 \includegraphics[scale=0.7]{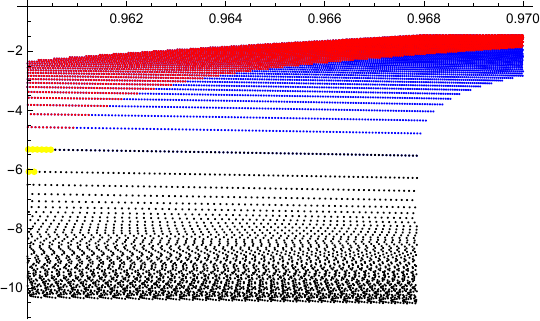}
~
 \includegraphics[scale=0.7]{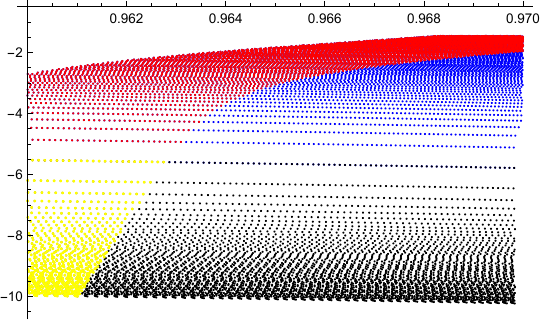}
\end{center}

 \baselineskip=15pt
\begin{quote}
{\bf Fig. 1} ~The  $(n_{\cR\cR},r)$ distributions for   $p=4,7,10,20$ in the panels from 
the upper left to the lower right respectively.  
Here blue (red) indicates the super-Planckian $\mu$ without (with) the reheating constraint, 
while black (yellow) shows sub-Planckian domains without (with) the reheating constraint.
\end{quote}

\baselineskip=18.5pt

The overall shapes of the viable regions in the spectral-tensor plane of Fig. 1 are determined by a combination of constraints. 
  The upper and lower boundary curves of the distributions come from the lower and upper bound of the number
   of inflationary e-folds.   A smaller interval of allowed values $N_*$ would lead to regions that are more narrow.  
 The flat plateaus of the upper regions of the distributions in Fig. 1 show that the {\sc Planck} bound  on $r$ 
   constrains these models for all $p$.  Future experiments like the Simons Observatory \cite{simons18}, 
    CMBS-4 \cite{cmbs4first, a19etal}  and LiteBird \cite{litebird22}  aim to reach $r \cong 0.001$ and will put more 
 pressure on some of these models if gravitational waves are not found. In particular 
  the often considered quartic model could be excluded and the same applies even more to the cubic hilltop model because 
  its lower bound of $r$ is higher than for the quartic model. For larger $p$ none of the upcoming CMB experiments 
  will be able to exclude these models based on their target values of $r$ in the range of a milli, even if the energy 
  scale $\mu$ is restricted to super-Planckian values. 
  
  The results in Fig. 1 also show that for low $p$ hilltop models improved constraints for the spectral index alone have the potential to 
significantly reduce the viable region in the spectral-tensor plane, while for higher $p$ such constraints will have less impact since
for larger $p$ the region in these graphs covers the whole range of the spectral index, at least in the tensor ratio range considered
here for sub-Planckian energy scales $\mu$. This pattern is stable when  $p$ is increased to $p > 20$ as far as the CMB constraints
 above are concerned.  Including the reheating constraint does not change  the viable range of $n_{\cR\cR}$ but does
 decrease  the regions dramatically by its effect on the tensor ratio. An improved accuracy of the 
spectral index $n_{\cR\cR}$ could again limit the viability of this class of models with a matter dominated reheating phase.
The combination of the results of the {\sc Planck} probe with the upcoming large scale survey data, such as Euclid and SKA, 
will therefore have a significant impact on the hilltop models. A discussion of the uncertainties of different combinations 
of these three experiments is for example given in ref. \cite{s18etal}. 

The $\mu$-stratifications of Fig. 1 show that in hilltop inflation the tensor-to-scalar ratio $r$ follows scaling relations 
in dependence of  $n_{\cR\cR}$ and that the scaling parameters vary as the energy scale $\mu$ is varied. 
This will be analyzed quantitatively  in later sections by extending and generalizing the strategy adopted
  in the context of multifield inflation in ref \cite{ls22}. Here we point to the fact that this energy dependence generalizes 
and refines  assumptions made in the literature that neglect a possible variation of the amplitude and the exponents in the scaling
 relation \cite{jyp22, g22etal}. 
 These results make explicit the difference between the strong small-field approximation 
 discussed earlier in this paper and the slow-roll approximation.  In the former the scaling exponents are 
 energy independent, while Fig. 1 illustrates the significant variation of these parameters over the viable parameter space.
As a result of this energy dependence the parameter spaces of the different hilltop models are 
constrained to a different degree, depending on $p$. The $(r, n_{\cR\cR})$ scaling relations of Fig. 1 
can be viewed as a consequence of the energy dependent scaling relations of the spectral 
tilt $\delta_n(N)$ and the tensor-ratio $r(N)$ as functions of the number of inflationary e-folds, which will be discussed in the 
next section. 

The $\mu$-stratification of the spectral-tensor plane also establishes correlations between 
some observables and the energy parameter that defines the deformation of the model.
  These are of different strength in that the correlation, for example, between the energy $\mu$ and the tensor ratio is 
quite strong, while the correlation between $\mu$ and the spectral index is much weaker.
 Other correlations, such as the anti-correlation between the number of inflationary e-folds $N_*$ and the reheating e-folds $N_\rmrh$
 depend on the model, becoming weaker as $p$ increases. The bounds on these e-folds are however stable and lead to upper 
 bounds of about forty and about sixty for the reheating e-folds and the inflationary e-folds, respectively.
 
\vskip .2truein

\section{Scaling distributions in hilltop inflation via $N_*$}

Scaling functions in dependence of the model parameters can be used as characteristics of the inflationary theory 
space, as discussed in the context of multifield theories in \cite{ls22}.  While this is a particularly interesting tool in 
multifield models with nontrivial potentials, even singlefield inflationary theories are usually not solvable analytically. 
Making sufficiently strong approximations to yield analytical methods leads to a loss of significant features. 
To gain a broader perspective of scaling in inflation we  analyze  the scaling behavior of hilltop inflation for both 
the spectral tilt  $\delta_n$ and the tensor ratio $r$ as functions of the number of e-folds.
We extend this analysis for a wide range of models and a large range of $\mu$-values that have been shown 
in the previous section to be viable (see e.g. 
 the distributions of the viable $\mu$-data in the $(n_{\cR\cR},r)$-plane for different $p$ shown in Fig. 1).
 
 In Fig. 2 we collect  the distributions for $\delta_n(N, \mu)$ as $p$ increases and $\mu$ is varied over its 
 viable range. The results show that $\delta_n$ scales with 
$N$ in dependence of the energy scale $\mu$ 
\beq
 \delta_n(N,\mu) ~=~ \frac{\a_{\delta_n}(\mu)}{N^{\b_{\delta_n}(\mu)}},
 \lleq{deltanN-scaling}
 where the existence of families of the energy dependent exponents extends the limit defined by
  the strong small-field approximation SF.III  with its energy independent scaling exponent.
  Fig. 2  shows again that the range of viable 
  e-folds is dramatically affected by the reheating constraint.
 \begin{center}
 \includegraphics[scale=0.7]{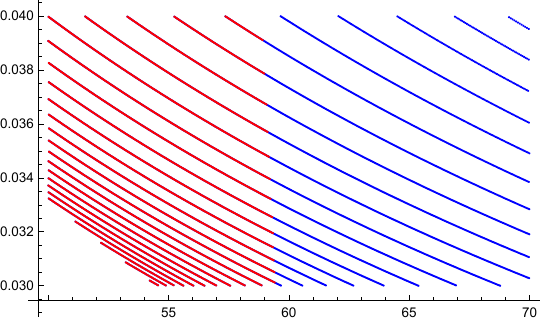}
~~ \includegraphics[scale=0.7]{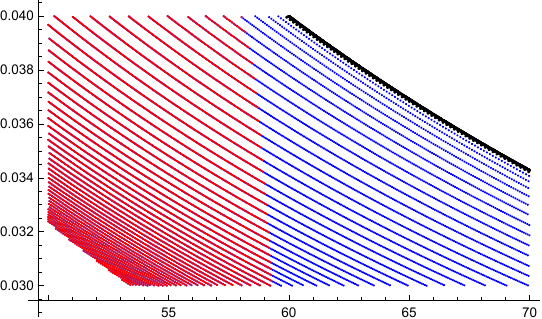}
 \includegraphics[scale=0.7]{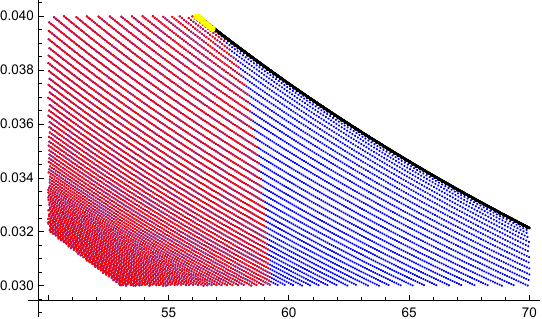}
~~  \includegraphics[scale=0.7]{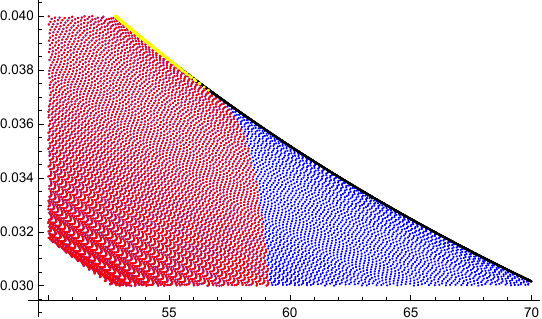}
 \end{center}
\begin{quote}
{\bf Fig. 2} ~Scaling distributions for $(\delta_n,N_*)$ in hilltop inflation models with $p=4,7,10,20$.
 The color scheme here is the same as in Fig. 1. 
\end{quote}

It was noted in the previous section that there is a weak correlation between the energy scale $\mu$ and the 
number of inflationary e-folds $N_*$. In Fig. 2 this degeneracy can be seen explicitly for the different models 
and that the distribution does become more narrow as $p$ increases.

In Fig. 3 we show the scaling distributions of the tensor-to-scalar ratio $r(N,\mu)$ as $p$ and $\mu$ are varied
on a logarithmic scale. The resulting behavior 
\beq
 r(N,\mu) ~=~ \frac{\a_r(\mu)}{N^{\b_r(\mu)}}
 \lleq{rN-scaling}
 again shows an energy dependence of the exponents that goes beyond the small-field approximation.
 
\begin{center}
 \includegraphics[scale=0.7]{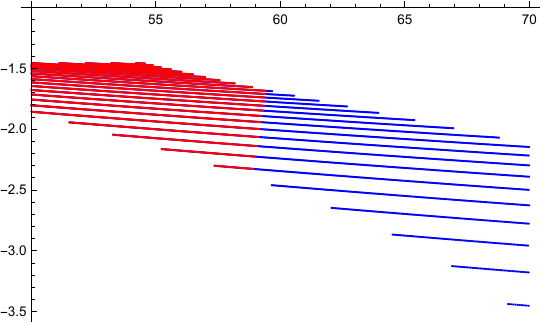}
~~  \includegraphics[scale=0.7]{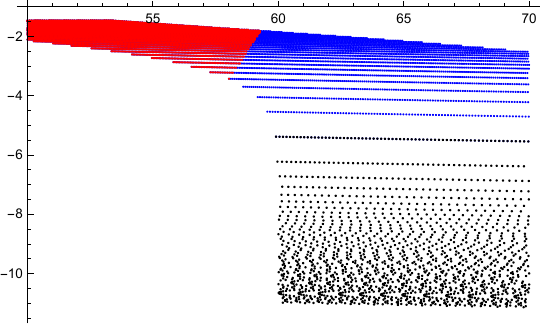}
 \includegraphics[scale=0.7]{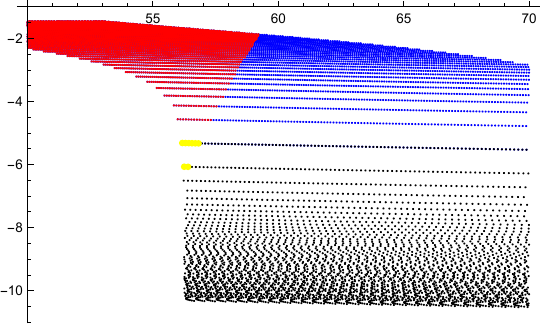}
~~ \includegraphics[scale=0.7]{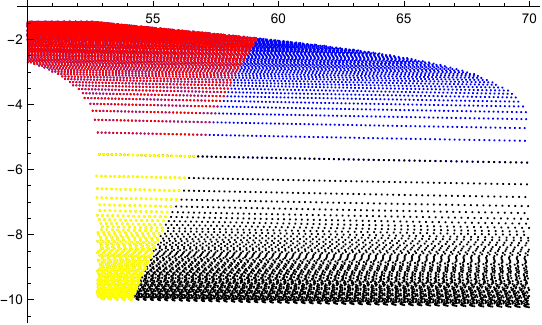}
 \end{center}
 \centerline{
{\bf Fig. 3} ~The tensor ratio distribution $(r,N)$ for $p=4,7,10,20$.
}

The gap at low $r$ and low $N$ in the graphs of Fig.3 is quite pronounced for small $p$ but becomes smaller 
as $p$ increases. This pattern continues for $p>20$. Furthermore the viable regions including the reheating
constraint increase in particular for the sub-Planckian range of $\mu$.

The two scaling relations (\ref{deltanN-scaling}) and (\ref{rN-scaling}) of Figs. 2 and 3 explain the scaling 
relations in Fig. 1, leading to a non-linear relation between the spectral tilt and the tensor ratio. 
They also explain the scaling behavior of the number of reheating 
e-folds and the reheating temperature as functions of the spectral tilt $\delta_n$ or the tensor ratio $r$.
This will be analyzed quantitatively in the next section.

\vskip .2truein
 
 \section{Scaling exponents in hilltop inflation}
 
 In this section we make the scaling behavior of the spectral tilt and the tensor ratio of the hilltop potentials more precise. 
  In  Figs. 1, 2 and 3  the scaling behavior of $r(\delta_n)$, $\delta_n(N)$ and $r(N)$ is shown for several hilltop 
  models in a systematic way for the full super-Planckian range of $\mu$-values as well as a partial range of sub-Planckian values.
 The resulting stratifications in these distributions suggest a scaling behavior of the form
 $\delta_n(N) = \a_\delta N^{-\b_\delta}$ and $r(N) = \a_r N^{-\b_r}$, 
  with exponents $\b$ and amplitudes $\a$ that depend on the energy scale $\mu$. 
  Hence we obtain for a given model a family of exponents for each of these observables. This 
  is qualitatively similar to the multifield behavior considered in \cite{ls22}.
 In the present discussion we will determine  the resulting exponents $\b$ explicitly in order to 
 be able to compare these characteristic numbers of hilltop inflation with those of other models and place 
 this class in the inflationary landscape.
   The results below can be compared  to the small $\mu$-limit of the small-field approximation prediction 
where both exponents are constants for a fixed model.  In the slow-roll approximation these families of exponents
 instead lead to a non-trivial  functional behavior.

 \subsection{Scaling exponents for the quartic hilltop model}
 
As noted above, the hilltop model at $p=4$ has been discussed extensively in the literature, 
see e.g. \cite{km95a, km95b, l96, bl05, b08etal, mrv13, l18, c19etal, kl19, d20, g20, hs21, c22}, 
and has been adopted as a bench-mark model by the CMB collaborations
 (see e.g. \cite{a15etal, a18etal, a19etal}). We therefore begin our analysis with this model
  for the most general range of $\mu$. In the previous section we have seen that 
   this model leads to viable  initial values compatible with the CMB results 
  only for super-Planckian $\mu$ values in a fairly narrow range. 
 As a consequence parts of the analyses of some of the earlier references are no longer applicable.

The distribution in the spectral-tensor plane is shown in  the previous section in Fig. 1, both with and without the
 reheating constraint. In the present discussion we focus on the scaling behavior of the observables as functions of $N$.
 Fig. 2 illustrates again that the spectral tilt $\delta_n = 1-n_{\cR\cR}$ of the spectral index away 
  from the Harrison-Zeldovich spectrum $n_{\cR\cR}=1$ shows for the quartic hilltop model 
   scaling behavior in dependence of the inflationary e-fold number $N_*$ as indicated above
  in eq.  (\ref{deltanN-scaling}).  In Fig. 4 we make this scaling relation more explicit for several 
  curves describing $\delta_n(N)$ for the hilltop model at $p=4$.
  \begin{center}
\includegraphics[scale=0.5]{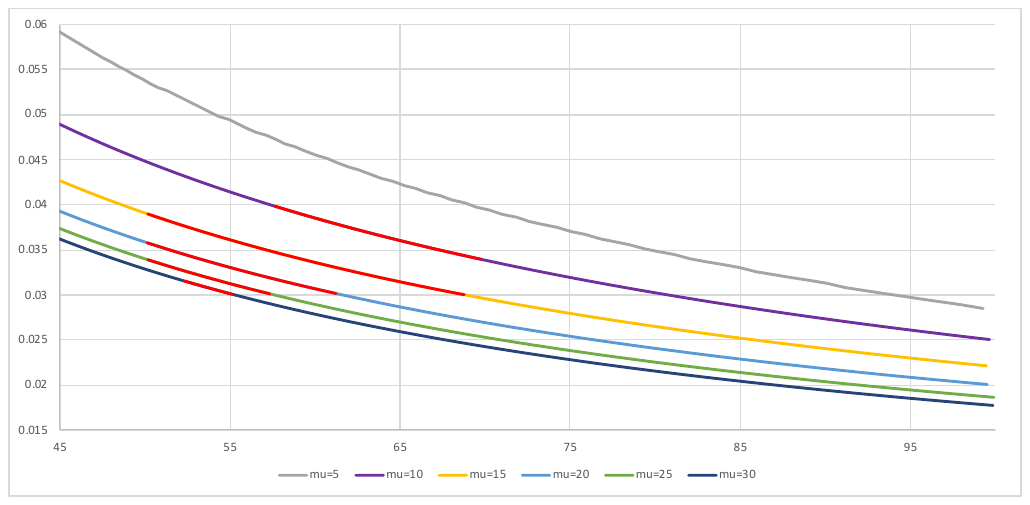}
  \end{center}
  \baselineskip=15pt
  \begin{quote}
  {\bf Fig. 4} ~Scaling behavior of the spectral tilt  $\delta_n(N,\mu)$ for the selection of $\mu$ of Table 1. Here the red sections 
  along the curves cover the range that is compatible with the CMB and the $e-$fold constraints.
\end{quote}

\baselineskip=19.5pt
  
  The curves in Fig. 4 lead to the exponents collected in Table 1.
   \begin{center}
  \begin{tabular}{l | c c c c c c}
  $\mu$                             &5             &10        &15          &20         &25     &30  \\
  \hline
  $\b_{\delta_n}(\mu)$  &0.924      &0.838   &0.833      &0.86     &0.886 &0.906  \tabroom \\
 \hline
  \end{tabular}
  \end{center}
  \vskip .05truein
  \baselineskip=15pt
  \begin{quote}
   {\bf Table 1.} ~Characteristic exponents $\b_{\delta_n}(\mu)$ for the spectral tilt $\delta_n(N)$ in the quartic hilltop model.
   \end{quote}

\baselineskip=19.5pt 

 The results of Fig. 4 and Table 1 show an energy dependent scaling exponent $\b_{\delta_n}(\mu)$, which for 
small $\mu$ approaches the limiting value  of the SF.III approximation, given by $\b_{\delta_n}=1$,
while the amplitude is in this limit given  by $2(p-1)/(p-2) = 3$.
 Roughly speaking, the slow-roll approximation 
  leads to a scaling behavior for $p=4$ hilltop of the form  $\delta_n(N) \sim N^{-0.8*}$ in the $\mu$-range 
  indicated in the above Table 1.

For the tensor ratio $r$ the first panel in Fig. 3 shows a similar power law scaling behavior 
$r(N) ~=~ \a_r N^{-\b_r}$  for the quartic hilltop model and a similar analysis leads to scaling functions that for a selection 
of $\mu$-scales are illustrated in Fig. 5.
\begin{center}
\includegraphics[scale=0.45]{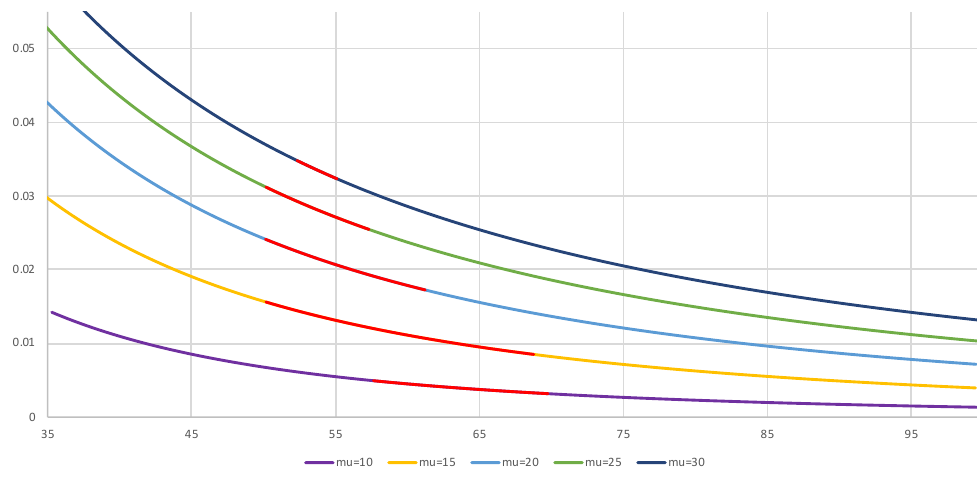}
\end{center}
\baselineskip=15pt
\begin{quote}
{\bf Fig. 5.} ~Scaling behavior of the tensor ratio $r(N)$ curves for quartic hilltop inflation.
 \end{quote}
 
 \baselineskip=19.5pt

The  scaling exponents  $\b_r$ associated to the curves of $r(N)$ in Fig. 5 lead to the results in Table 2.
\begin{center}
\begin{tabular}{l | c c c c c  c}
$\mu/M_\rmPl$     &5      &10       &15      &20         &25  &30 \\
\hline

$\b_r(\mu)$                  &2.748  &2.268    &1.897   &1.667   &1.532  &1.446 \tabroom \\

\hline
\end{tabular}
\end{center}
\baselineskip=15pt
\vskip .05truein
\centerline{
{\bf Table 2.} The scaling exponents of the tensor ratio $r(N)$ at $p=4$.
}

\baselineskip=19.5pt

This can be compared to the analytical tensor ratio scaling behavior obtained in  the  SF.III approximation.
 In this approximation the scaling exponent $\b_r$ is independent of $\mu$ and 
 its value is given by $\b_r^{\rmIII} = 2(p-1)/(p-2) = 3$.
The above results show the $\mu$-dependence of  the scaling exponents $\b_r(\mu)$ and
that for small $\mu$ they approach the value of the strong small field approximation SF.III.

 With the scaling relations $\delta_n(N)$ and $r(N)$ determined above we can also eliminate $N$ in the second 
by using the tilt scaling to obtain a scaling relation of the form $r=r(\delta_n)$. The resulting exponents 
$\b_r$ and $\b_{\delta_n}$ can be compared with the 
analytical scaling relation determined in the SF.II approximation in eq. (\ref{SFII-r-vs-delta}).
 Alternatively, one can analyze the distributions of Fig. 1 directly in the same way as done above for the 
$N$-scaling functions.  This places $r(\delta_n)$ directly in spectral-tensor plane and with a measurement 
of $r$ would allow tests not only of the model but also of the parameter space spanned by $\mu$.

The post-inflationary evolution constraint is illustrated in this paper with a matter dominated reheating phase.
The implications of the reheating constraint for the different models are shown in 
the first three figures for both the super-Planckian and the sub-Planckian energy scales $\mu$. For 
hilltop inflation with $p=4$ the reheating e-fold range is with our run-parameters  roughly $\Delta N_\rmrh=40$ 
while the reheating temperature $T_\rmrh$ is roughly between $10^3\rmGeV$ 
 and $10^{15}\rmGeV$.  
 
\vskip .2truein

\subsection{Behavior of  hilltop inflation at $p=10$}

 For $p\geq 5$ the range of viable energies $\mu$ includes the sub-Planckian regime. In the present section
we briefly summarize here the scaling results at $p=10$, which are obtained in the same way as described in detail
in the previous subsection for $p=4$.   For this model the sub-Planckian values for the energy scale $\mu$ lead to 
 a much deeper range for the tensor ratio.
 
 The distribution $(\delta_n,N)$ for all viable initial values at $p=10$ takes the form shown in Fig. 2.
 Similar to the case $p=4$, the  distribution at $p=10$ is again formed by a set of curves (that are described 
 by a scaling relation in dependence of $\mu$ of the type in eq. (\ref{rN-scaling}). 
 The scaling curves for a selection of energy scales $\mu$ are collected in Fig. 6.
 \begin{center}
 \includegraphics[scale=0.55]{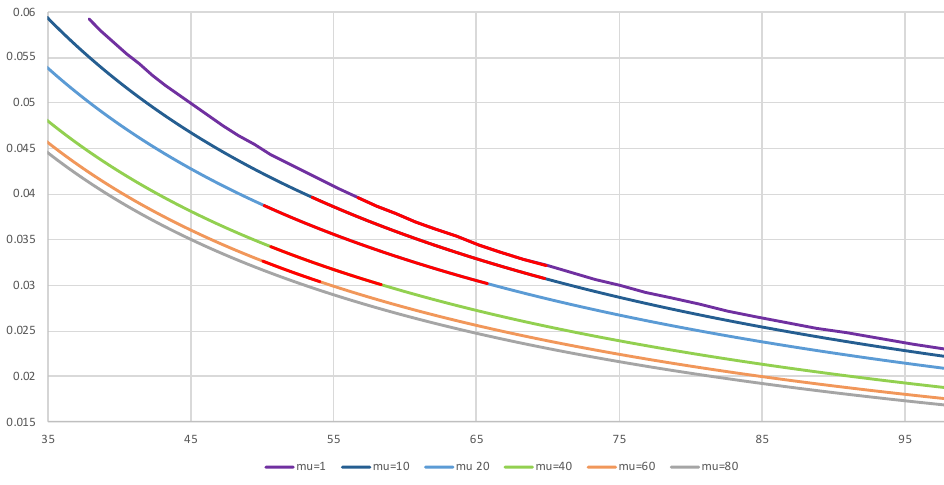} 
\end{center}
\centerline{{\bf Fig. 6} ~Scaling behavior of $\delta_n(N)$ for $p=10$ hilltop inflation.}

 The scaling  exponents for the curves in Fig. 6 can be found in Table 3. The values obtained can again be 
compared to those of the small-field approximation, where the amplitude evaluates for $p=10$ 
to $\a_{10}=9/4$.

 \begin{center}
 \begin{tabular}{l | c c c c c c}
 $\mu/M_\rmPl$                               &1         &20        &40         &60        &80       \\
 \hline
 $\b_{\delta_n}(\mu)$    &0.998  &0.922   &0.916    &0.934    &0.947   \tabroom \\
  \hline
 \end{tabular}
 \end{center}
 \centerline{
 {\bf Table 3.} ~Scaling exponents $\b_p$ for $\delta_n$ in hilltop inflation at $p=10$. 
 }
    
We consider again the behavior of the tensor-to-scalar ratio $r$, for which 
 Fig. 7 shows the dependence on the inflationary number of e-folds for $r(N_*)$ for 
several energy values $\mu$ extracted from the distribution of $\mu$-family curves in Fig. 3.
\begin{center}
\includegraphics[scale=0.55]{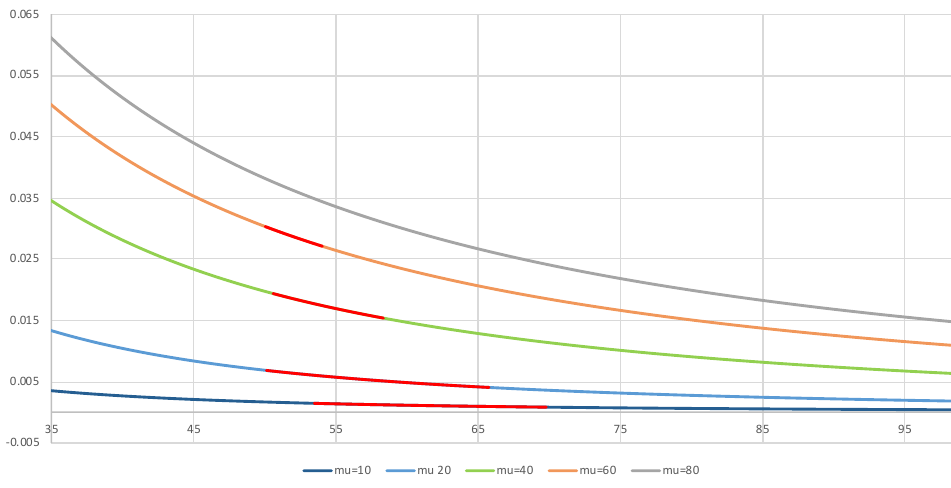}
\end{center}
\centerline{{\bf Fig. 7} ~Scaling behavior of $r(N)$ at $p=10$.}

The scaling curves in Fig. 7 lead to the  scaling exponents compiled in Table 4. 
These results can again be compared to the analytical tensor ratio scaling behavior that can be obtained in 
 the SF.III approximation. The latter predicts that $\b_r^\rmIII = 2.25$.

\begin{center}
\begin{tabular}{l | c c c c c}
$\mu/M_\rmPl$  &1      &20          &40       &60  &80 \\
\hline
$\b_r(\mu)$       &2.24  &1.92    &1.62   &1.46   &1.37 \tabroom \\
\hline
\end{tabular}
\end{center}
\vskip .05truein
\centerline{
{\bf Table 4.} The scaling exponents of $r(N)$ for $p=10$ hilltop inflation.
}

 The reheating parameters $N_\rmrh$ and $T_\rmrh$ for the $p=10$ hilltop model are comparable to the 
 values for the quartic hilltop model.

As the exponent $p$ of the hilltop models increases the pattern of the distributions in Figure 1 is stable in that 
the spectral-tensor plane region that is viable under CMB + e-fold constraints remains saturated as for the $p=20$ 
model.  Furthermore,  the range of the scaling exponents decreases and remains within the viable region for larger 
ranges for the energy scale $\mu$. 

The above results for the hilltop models at $p=4$ and $p=10$ already indicate that these two inflationary scenarios 
can be characterized by the ranges they define for the scaling exponents of the spectral tilt and the tensor ratio. 
This will be discussed in more detail in section 9, after the analysis of the hilltop-squared models.

\subsection{Implications} 

The energy dependence of the $N-$scaling behavior of any observable 
\beq
  \cO(N) ~=~ \a_\cO N^{-\b_\cO},
  \lleq{N-scaling-of-observable}
 can be used to determine how the uncertainties of parameters 
  describing the inflationary and reheating phases are affected by varying $\mu$. 
  Denoting these parameters by $\cP$, for example the e-folds of the inflationary and reheating phases,  
  and the reheating temperature, their uncertainty is 
  determined by the uncertainties $\Delta \cO$  associated with either the measured values 
 of $\cO$, or the uncertainties of fiducial values predicted for future experiments.
    By inverting (\ref{N-scaling-of-observable}) the induced uncertainty of the inflationary number of 
 e-folds is determined by
 \beq
  \Delta N~=~ - \frac{\a_\cO^{1/\b_\cO}}{\b_\cO}  \frac{\Delta \cO}{\cO^{1+\frac{1}{\b_\cO} }}
  \eeq
  which depends on $\mu$ via the energy dependence of the  amplitude $\a_\cO(\mu)$ and the 
  exponent $\b_\cO(\mu)$.
This in turn can be used to trace the implications of the uncertainties of $\cO$  for $\Delta T_\rmrh$. To do so it is useful 
to make the dependence of $T_\rmrh$ on the number of inflationary e-folds explicit by writing 
 eq.  (\ref{Trh-for-general-f}) for the reheating temperature via eq. (\ref{Nrh-for-general-f}) for the number of reheating 
e-folds  as
\beq
 T_\rmrh ~=~ \k F(\phi_*) \exp\left( \frac{3(1+w_\rmrh)}{(1-3w_\rmrh} N_*\right),
 \eeq
 where 
  \beq
   F(\phi_*) ~=~ \left( \frac{f_*^{3w_\rmrh}}{(f'_*)^{1+3w_\rmrh}}\right)^{1/(1-3w_\rmrh)}
   \eeq
   and $\k$ collects all the remaining terms that are independent of $\phi_*$ and $N_*$. With these ingredients 
  we obtain
  \beq
   \frac{\Delta T_\rmrh}{T_\rmrh} ~=~  \left(\frac{3(1+w_\rmrh)}{1-3w_\rmrh} ~+~ \frac{F'_*}{F_*}\right) \Delta N_*.
   \eeq
 where here the dimensionless derivative is relative to the number of e-folds $F' = dF/dN$.  
 The energy dependence of the scaling amplitude $\a_\cO(\mu)$ and the exponent $\b_\cO(\mu)$ therefore leads 
  to a $\mu$-dependence of these uncertainties. Using for example the estimates of the CMB-S4 
  and the Simons collaborations \cite{cmbs4first, a19etal, simons18}  this leads to uncertainties induced by 
  the spectral index and the tensor ratio.
  
 \vskip .3truein

\section{Bifurcation among $p$-families in hilltop inflation}

In the previous sections we have analyzed the behavior of the spectral tilt $\delta_n$ and the 
tensor-to-scalar ratio $r$ of hilltop models in dependence of model parameters $p$ and  $\mu$, 
as well as the number of e-folds $N$, leading to functions on the 
theory space. Our strategy has been to consider a wide range of models and determine 
the energy ranges that are compatible with  the constraints provided by the CMB data and the e-fold constraints.
This leads to energy dependent scaling exponents that we interpet as diagnostic tools to characterize 
inflationary models. 

An alternative view of the class hilltop models can be obtained by varying the models systematically 
by letting $p$ run while keeping  $\mu$ and $N_*$ fixed, leading to a $p$-family. 
By subsequently varying $\mu$ we obtain a set of $p$-families in the spectral-tensor plane,
as shown in Fig. 8.  The strategy here is to determine $\phi_e$ in the slow-roll approximation
as the solution of the degree $p$-constraint of eq.  (\ref{phie-via-epsV})  
and  to determine $\phi_*(N,p)$ via eq.  (\ref{phistar-via-N}).
This type of analysis gives further insight into the differences between the hilltop models and it also illustrates 
the energy dependence of the observables described earlier in this paper in a different way.

\subsection{The  bifurcation phenomenon for  $p$-families}
 
 Naively, one might have expected that as $\mu$ increases the $p$-family curve 
 becomes ever steeper in the spectral-tensor plane,
perhaps approaching the vertical line. This turns out not to be the case.
Instead, when changing the $\mu$-parameter we find that at some ``critical value" $\mu_c$ 
 a bifurcation arises in the plane spanned by the spectral index 
 and the tensor ratio. 
 For low $\mu$ the trajectories begin at low $n_{\cR\cR}$ for low $p$-values, with increasing $n_{\cR\cR}$
 as $p$ increases. Beyond the critical value of $\mu$ the $p$-families begin for small $p$ at large $n_{\cR\cR}$, with 
 decreasing $n_{\cR\cR}$ as $p$ increases.  This is illustrated in Fig. 8, which also contains an overlay of the 
 forecasts of the Simons Observatory to illustrate how different models are affected by experimental constraints.
  The precise location of the bifurcation in the plane depends on the choice of the e-folds. 
  In a first approximation this location is clear from the approximation of the spectral index in the 
  form of eq. (\ref{specind-triple-approx}), which ignores the $\phi_e$ dependence of the e-folds. 
  In this approximation the large $p$-limit 
  approaches the spectral index $n_{\cR\cR} = 1- 2/N_*$.
\begin{center}
\includegraphics[scale=0.25]{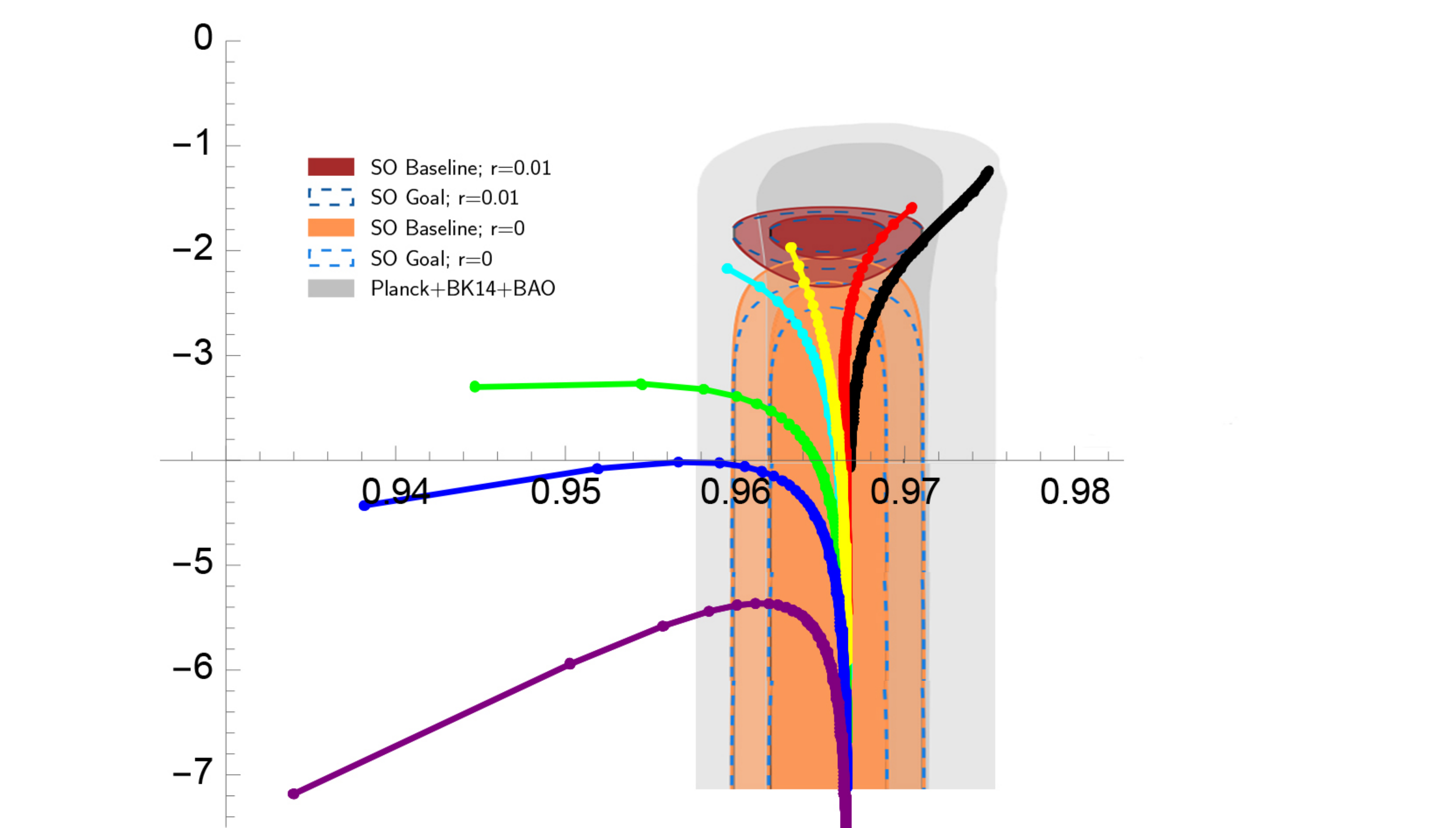}
\end{center}
\baselineskip=15pt
\begin{quote}
 {\bf Fig. 8}~ A selection of $p$-families in the $(n_{\cR\cR}, r)$-plane for different energy scales  $\mu/M_\rmPl$ 
 values   between one (purple, low left corner) and one-hundred (in black,  upper right corner). 
 The forecasts here are by the Simons Observatory collaboration \cite{simons18} for different scenarios.
 \end{quote}
 
 \baselineskip=19.5pt

The structure of the $p-$families in Fig. 8  illustrates in a different way the results of  the scan analysis earlier in this paper, 
in particular  the fact that for any super-Planckian energy scale $\mu$ there are hilltop models that are compatible with the CMB and 
lead to sufficient inflation.
It also illustrates that for low $p$-models there is a cut-off in the super-Planckian regime below which 
no viable models can be found. This includes in particular the quartic model for which 
sub-Planckian energy is ruled out.

\subsection{Small-field structure of $p$-families}

It is clear that the bifurcation of the $p$-families in the spectral-tensor plane cannot be understood
 in the approximation of hilltop inflation in which the dependence of the number of e-folds on the field 
 $\phi_e$ at the end of inflation is neglected. 
This approximation leads to the spectral index (\ref{specind-triple-approx}) and the 
tensor-ratio (\ref{tensor-ratio-triple-approx}) considered in the early 
literature \cite{km95a, km95b, l96}. The reason why the $\phi_e$-terms have often  been neglected 
is because the focus was
 on the low range of the energy scale $\mu$, in which case  the $N$-independent terms in the denominators of 
 the small-field approximation are small and it is justified to neglect them.
An unusual feature of this procedure is that the spectral index does not depend on the energy scale. While it is usually the 
case that $n_{\cR\cR}$ does not depend on the overall energy scale $\L$, it does in general depend 
on the remaining parameters of the theory, and we see here 
that in  the less severe SF.II  approximation this is also   the case for hilltop inflation.
 
 To illustrate that the SF.II approximation does capture the bifurcation we consider in Fig. 9
 several $p$-family curves in this approximation for several $\mu$-values  between 1 and 200 $M_\rmPl$.
 This graph is truncated compared to the numerical slow-roll graph in Fig. 8 because 
 of the $p$-dependent consistency constraints on the energy scale $\mu$ in this approximation.
 Nevertheless, it does show that the bifurcation  phenomenon  can be captured by the SF.II 
 approximation even though the  detailed numerics of the trajectories  is of course different in pure slow-roll as 
 compared to the SF.II approximation.
  \begin{center}
    \includegraphics[scale=0.6]{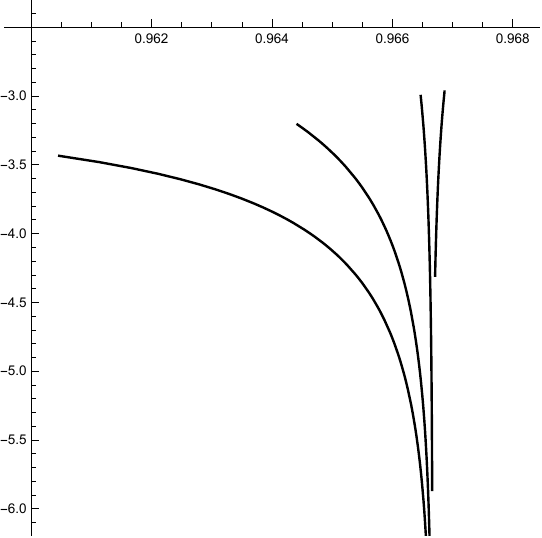}
 \end{center}
 \baselineskip=15pt
\begin{quote}
{\bf Fig. 9}~ An illustration of the SF.II approximation of the bifurcation phenomenon in hilltop inflation for 
energy scales between 5$M_\rmPl$ and 100 $M_\rmPl$ with $p$ running for $p>p_\mu$, where $p_\mu$ 
is the consistency bound.
\end{quote}

\baselineskip=19.5pt

\vskip.3truein

\section{Hilltop-squared inflation}

We have noted already that hilltop inflation $V_{p,n}$ as defined in eq. (\ref{hi+hsi}) with $n=1$  does not contain a minimum valley in which the 
inflaton can  oscillate during reheating. While this is in principle easy to fix with a small deformation of the 
potential close to its zero,  this feature has motivated the consideration of more global changes of the hilltop models,
such as hilltop-squared inflation (HSI), with potentials of the type considered in eq. (\ref{hi+hsi}) 
with $n=2$ \cite{dpb16}. 
Higher exponents $n$ have been considered in \cite{kl19, hs21, lr22, g23etal}.

In the present section we extend our analysis of the scaling behavior of the spectral tilt and the tensor ratio 
considered above for hilltop inflation to potentials of the type
\beq
 V_{p,2}~=~ \L^4 \left(1-\left(\frac{\phi}{\mu}\right)^p\right)^2.
 \lleq{hsi-models}
 We focus in particular on the phenomenological aspects and  the scaling behavior of these models.
 We find that the scaling results for the hilltop-squared models differ from those of the hilltop models, but that they follow 
 a similar overall pattern.

\subsection{Observables in hilltop-squared inflation}

The spectral index of the scalar perturbations for the hilltop-squared inflation models (\ref{hsi-models})  is given by
\beq
n_{\cR\cR}^\rmHSI  ~=~ 1 - \frac{4p\frac{M_\rmPl^2}{\mu^2}}{\left(1-\left(\frac{\phi}{\mu}\right)^p\right)^2}\left(\frac{\phi}{\mu}\right)^{p-2}
    \left[  (p-1)   + (p+1)  \left(\frac{\phi}{\mu} \right)^p  \right]
 \eeq
and the tensor ratio $r^\rmHSI$ relates to the hilltop tensor ratio as
\beq
 r^\rmHSI(\phi) ~=~  32p^2    \frac{M_\rmPl^2}{\mu^2}
  \frac{ \left(\frac{\phi}{\mu}\right)^{2(p-1)} }{\left(1-\left(\frac{\phi}{\mu}\right)^p\right)^2}
   ~=~ 4r.
  \eeq
 The inflaton field at the end of inflation $\phi_e$  is determined in HSI$_p$ by 
 \beq
  x_e^p ~+~ \sqrt{2}p \frac{M_\rmPl}{\mu} x_e^{p-1} ~-~ 1 ~=~ 0,
  \eeq
   where $x_e= \phi_e/\mu$.  The number of e-folds $N_*^\rmHSI$ is related   to that of hilltop inflation as
  \beq
 N_*^\rmHSI ~=~ \frac{1}{2} N_*^\rmHI.
 \eeq
 The reheating e-fold number $N_\rmrh$ and the reheating temperature $T_\rmrh$ can be obtained 
 from the general formulae given above in (\ref{Nrh-for-general-f}) and (\ref{Trh-for-general-f}) via 
 the dimensionless hilltop-squared potential.
 
 \subsection{Distribution of $(r,n_{\cR\cR})$ in hilltop-squared inflation}

In the present subsection we consider a sample of HSI$_p$ models for the same range of
 $p$-values considered above for hilltop inflation and with the same CMB and reheating constraints so as to be 
  consistent with the {\sc Planck} results for the observables. The  number of 
e-folds is again in the canonical interval $[50, 70]$.
 Fig. 10  shows the distribution in the spectral-tensor plane with the same color code that was 
 used in Fig. 1 for the hilltop models.
\begin{center}
\includegraphics[scale=0.65]{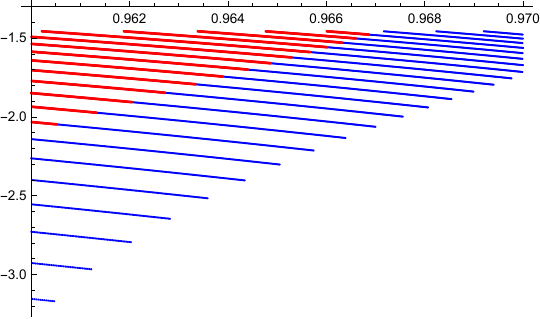}
~~\includegraphics[scale=0.65]{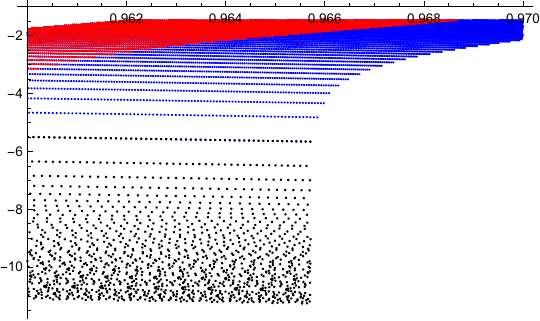}

\includegraphics[scale=0.65]{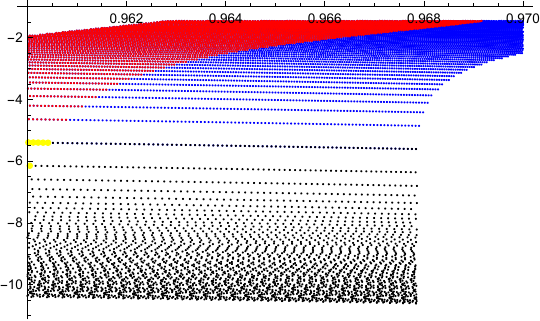}
~~\includegraphics[scale=0.65]{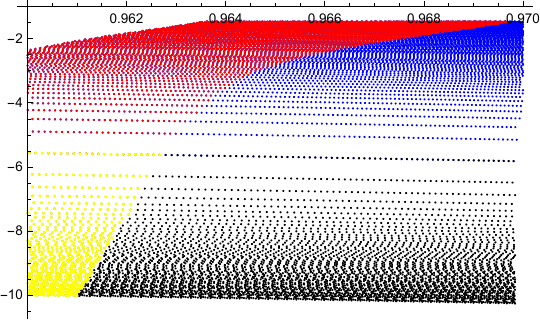}
\end{center}
\baselineskip=15pt
\begin{quote}
{\bf Fig. 10}~ The spectral-tensor distributions for hilltop-squared models with $p=4,7, 10, 20$ from 
top left to bottom right.
For $p=4,7$ the reheating constraint rules out sub-Planckian $\mu$.
\end{quote}

\baselineskip=19.5pt

These results imply that hilltop-squared inflation is not small-field inflation in general in the sense 
that for some $p$-models sub-Planckian field values are not compatible with the CMB constraints. 
This is similar to the behavior of the hilltop models discussed above. 
The panel in Fig. 10 for hilltop-squared at $p=4$ shows in particular that the allowed range in the spectral-tensor 
plane  is quite small and that there are in particular no
sub-Planckian energies $\mu$ that lead to viable initial values. One can check that this implies that 
the inflaton values $\phi_*$ for this model are all super-Planckian and hence hilltop-squared inflation at 
$p=4$  is not  small-field inflation. The same holds for the $p=3$ model. 

\subsection{Scaling distributions for hilltop-squared observables via $N$}

The behavior of the spectral tilt  $\delta_n$ is of interest both from a theoretical and 
a practical perspective, as discussed above in the context of the hilltop models. 
 Fig. 11 shows the scaling distributions for the spectral tilt $\delta_n$ for different $p$-models in hilltop-squared inflation. 
 Like in hilltop inflation these distributions show a weak correlation between the energy scales $\mu$ and the number of 
 inflationary e-folds. 
 \begin{center}
 \includegraphics[scale=0.6]{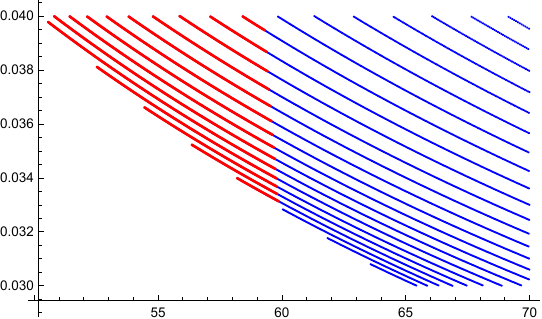} 
~~ \includegraphics[scale=0.6]{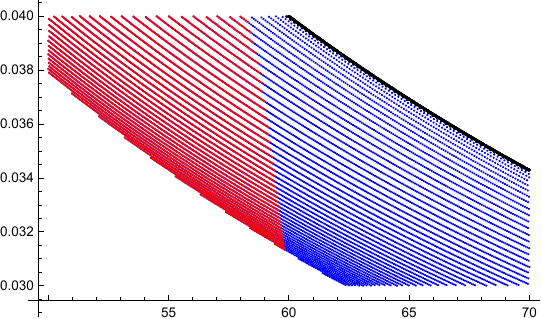} 

 \includegraphics[scale=0.6]{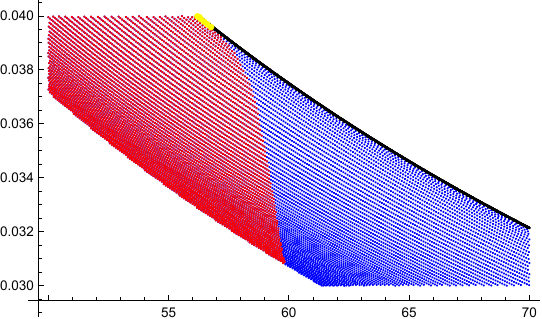} 
~~ \includegraphics[scale=0.6]{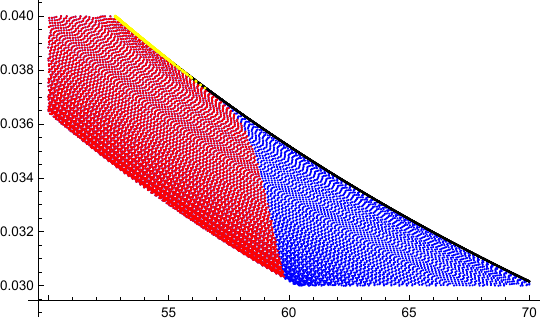} 
\end{center}
\baselineskip=15pt
\begin{quote}
{\bf Fig. 11} ~$\delta_n(N)$ distribution for $p=4,7,10,20$.  The color coding is the same as
 in the hilltop inflation distributions. For $p=4$ there are no contributions from 
sub-Planckian $\mu$, hence there are no black and yellow regions in the corresponding panel.
\end{quote}

\baselineskip=19.5pt
 
\subsection{$r(N)$ scaling in hilltop-squared inflation}

The behavior of the tensor-to-scalar ratio $r$ as a function of the number of inflationary e-folds $N_*$ 
in hilltop-squared inflation is similar to that of hilltop inflation. Fig. 12  illustrates the $(r,N)$ scaling relations 
for several hilltop models. As in the case of the hilltop models the viable region for $p=4$ is much smaller 
than that of the higher $p$ models.

\begin{center}
\includegraphics[scale=0.6]{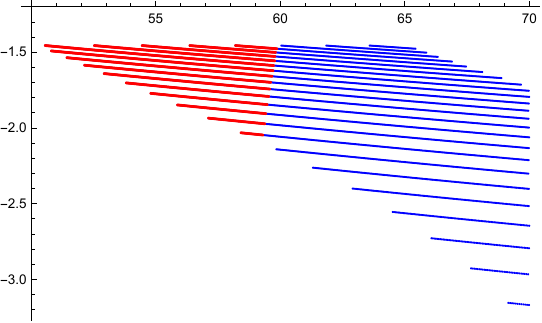}
~~\includegraphics[scale=0.6]{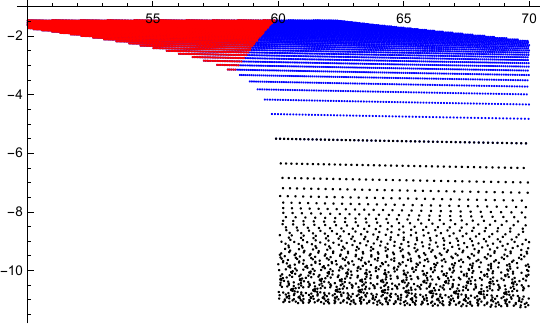}
\end{center}
\begin{center}
\includegraphics[scale=0.6]{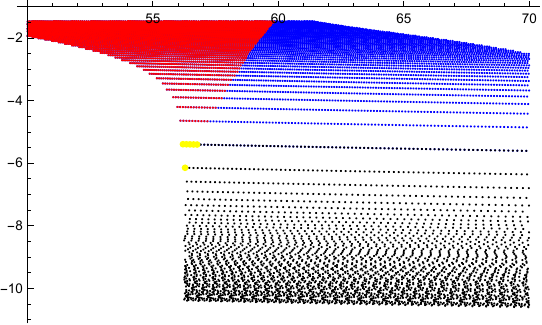}
~~\includegraphics[scale=0.6]{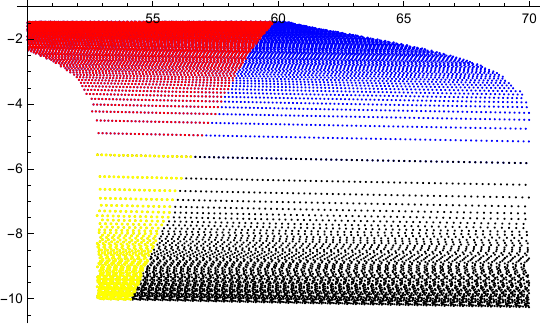}
\end{center}
\centerline{
{\bf Fig. 12} ~ Distributions of $(r,N)$ for $p=4,7,10,20$.}

\vskip .3truein

\subsection{Scaling exponents in hilltop-squared inflation}

The  analysis of the scaling behavior in hilltop-squared inflation follows the same strategy as in  hilltop inflation.
The structure of the two classes of models is similar although the numerical results differ. 
We have analyzed a wide range of models between $p=3$ and $p=100$, based 
on the distributions discussed in the previous sections, and extensions thereof, but 
here we  focus on the $p=4$ hilltop-squared model to illustrate the results.
The distribution in the spectral-tensor plane shown in Fig. 12  establishes the regions that are viable given 
the CMB inflationary constraints with or without the reheating constraint.  In the following we focus
 on the scaling exponents of the tensor ratio in dependence of the number of e-folds, shown for several models in 
Fig. 12 as a function of the energy scale $\mu$. Some aspects of this model other than scaling have also been discussed in 
\cite{hs21}.
 In Fig. 13 we select several curves from the distribution shown in the $p=4$ panel of Fig. 12.
\begin{center}
\includegraphics[scale=0.6]{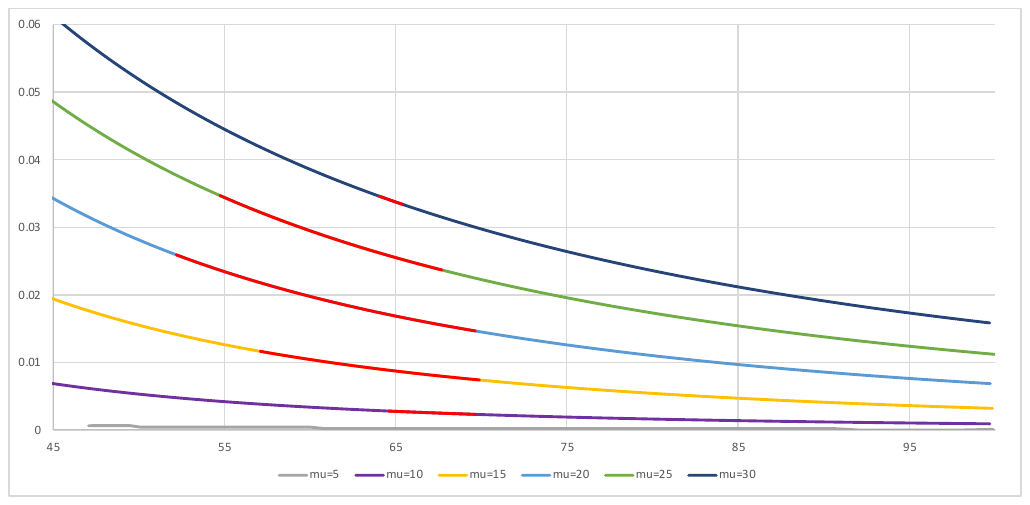}
\end{center}
\baselineskip=15pt
\begin{quote}
{\bf Fig. 13}~ A selection of $r(N)$ curves for hilltop-squared inflation at $p=4$.
\end{quote}

\baselineskip=19.5pt

The scaling exponents for the curves in Fig. 13 lead to the results listed in Table 5.
\begin{center}
\begin{tabular}{l | c  c c c c c}

$\mu$             &5          &10         &20              &25       &30 \\
\hline
$\b_r(\mu)$            &2.859   &2.546   &1.958   &1.798   &1.687  \tabroom \\

\hline
\end{tabular}
\end{center}
\centerline{
{\bf Table 5.} ~Scaling exponents $\b_r(\mu)$ for at $p=4$ HSI.
}
 
 For hilltop-squared models the range of the scaling exponents becomes more narrow as $p$ increases
because the upper boundary decreases and the lower boundary changes only slightly.

\vskip .3truein

\section{Scaling exponents as model characteristics}

In the previous sections we have determined the scaling exponents of the two main observables  
$n_{\cR\cR}, r$ as functions of the number of inflationary e-folds in both hilltop and hilltop-squared inflation models. 
These power law scaling exponents depend on the energy parameter $\mu$ of  the models, 
leading for a given model to functions $\b_\delta(\mu)$ and $\b_r(\mu)$.
More generally, we can think of observables $\cO(N)$ and their associated exponents $\b_\cO(\mu)$ as functions
on  the model parameter space. These energy scales are constrained, depending 
on the conditions imposed on the observables.  If one is interested in what present and future experiments can observe
then these constraints include those obtained from CMB and LSS data, assumptions about the number of inflationary 
e-folds $N_*$  and, in principle, assumptions about the reheating period and other constraints. 
Given such a set of $\mu$-dependent scaling exponents for an observable $\cO$ it is natural to ask how they 
characterize different models. The discussion above suggests that this can be achieved in different ways.
An immediate comparison of models can be obtained by comparing the 
ranges of the exponents for different models, leading to what we will call ``characteristic intervals". 
A second way is to focus on the functional form of the scaling exponents, leading to ``characteristic functions".

\subsection{Characteristic intervals}

One possibility to distinguish different inflationary models in an observable-based way is by comparing 
 the characteristic intervals obtained by imposing the CMB constraints and the number of inflationary e-folds. 
 The results above of the analyses of hilltop and hilltop-squared inflation  show that for both types of models the 
 strong small-field approximation  SF.III leads to an upper bound of the corresponding characteristic intervals. 
 For small $p$ this upper limit is quite different from the actual upper limit, but for larger $p$ the energy independent SF.III 
exponent $\b_r^\rmIII= 2(p-1)/(p-2)$ is quite close to the 
slow-roll upper bound of the tensor ratio scaling $r(N)$. 

In order to gain perspective it is useful to compare the hilltop and hilltop-squared results above with results obtained 
in other models.  In ref. \cite{ls22} a scaling analysis of the type considered in the present paper was constructed for 
two modular inflation models, theories that are based on softly broken groups $\rmSL(2,\IR)$, associated to 
a hyperbolic target space. 
In these models the potentials are invariant under discrete subgroups of the M\"obius group and the 
theories have two energy scales $(\L,\mu)$, much like the hilltop and hilltop-squared families. 
The observables $n_{\cR\cR}$ and $r$ can be viewed as functions of the number of e-folds and show 
scaling behavior that is also energy dependent. The resulting exponents again sweep out characteristic intervals
and hence can be compared with the results obtained in the present paper.  The fact that the $\b_r^\rmIII$ define the 
upper limit for the energy dependent exponents 
immediately implies that  the hilltop and hilltop-squared models define scaling classes that
 are distinct from the scaling classes defined by modular inflation models considered in
 \cite{ls22} because the characteristic intervals of hilltop and hilltop-squared inflation have no overlap with the models 
 considered there.  This shows that characteristic intervals provide a useful diagnostic for both singlefield and twofield inflation.

The comparison of the characteristic intervals of the two hilltop classes on the other hand 
shows that they are similar, with a somewhat 
higher lower bound of $\b_r(\mu)$ for hilltop-squared models as compared to hilltop models. 
Roughly speaking, the distribution 
of these values in the hilltop-squared class is more narrow than in hilltop inflation.

\subsection{Characteristic functions of scaling exponents}

A second defining feature that can be considered, once the distribution of the scaling exponents $\b_\cO(\mu)$ of 
an observable $\cO$ is known, is their functional form. Here we focus on the tensor-to-scalar ratio for hilltop and hilltop-squared 
inflation, similar to the analysis considered for modular inflation in ref.  \cite{ls22}.   In both HI and HSI inflation it is tempting 
to use as a guide for the functional fit the fact that for large $p$  the exponents $\b(\mu)$  are approximately flat 
because for large $p$ these models behave similar to the strong small-field approximation SF.III and in this approximation 
there is no $\mu$-dependence. This suggests a linear ansatz for the $\b(\mu)$ as a comparison to the horizontal line
determined by SF.III. It turns out however that  this is not a good approximation in general, as seen already earlier 
in the analyses above. 

 We make the functional $\mu$-dependence of the tensor scaling exponents  explicit in Fig. 14 for the classes of hilltop and 
 hilltop-squared inflation for six different models between $p=3$ and $p=100$ and an energy range $\mu \leq 150M_\rmPl$.
 The grey bands indicate the characteristic intervals of the approximate range of exponents for which the 
 models are  compatible with the CMB and e-fold constraints.
 For small $p$ these $\mu$ have a  lower bound that is super-Planckian, hence for these models the exponent functions
 leave this CMB region at smaller values of $\mu$ and again for larger $\mu$, depending on $p$.  For larger $p$ the $\mu$ 
 range reaches  down quite far  into the sub-Planckian regime and the functional form of scaling exponents $\b_r(\mu)$ 
 becomes flatter as the viable $\mu$-range expands.  It is only for the largest $p$ models considered here
 that the exponents remain completely within the characteristic band in the whole range of $\mu$
  considered in the graphs. 
  
The behavior of the $r(N)$-exponents in Fig. 14 for hilltop and hilltop-squared models shows that for the complete 
range of $\mu$-values, irrespective of the precise observational constraints, the functions $\b_r(\mu)$ can be described 
by higher degree polynomials for low $p$, with the degrees decreasing as the model exponents $p$ increase. These 
polynomial fits for the $\b_r(\mu)$ thus provide intrinsic characteristic functions of the models. In  the parameter 
ranges that are viable relative to the CMB and e-fold constraints the functions $\b_r(\mu)$ become simpler in the sense that 
fewer parameters suffice for a good description.

 \begin{center}
\includegraphics[scale=0.8]{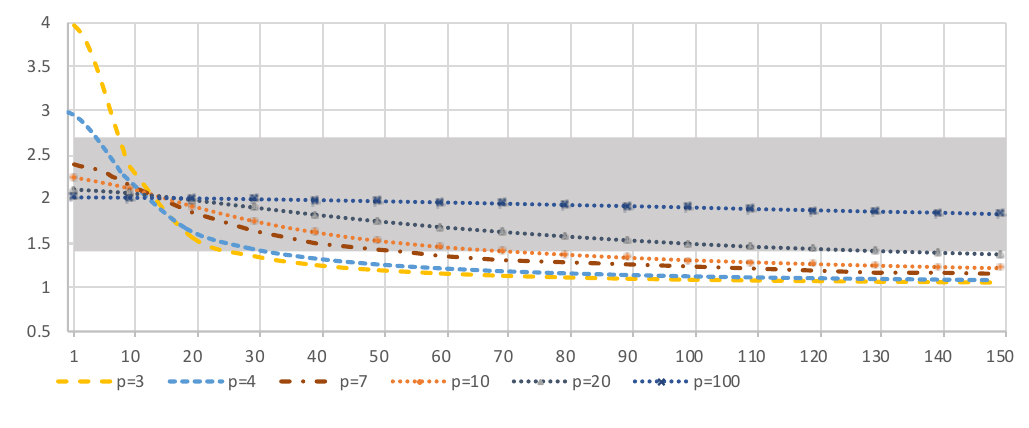}

\includegraphics[scale=0.8]{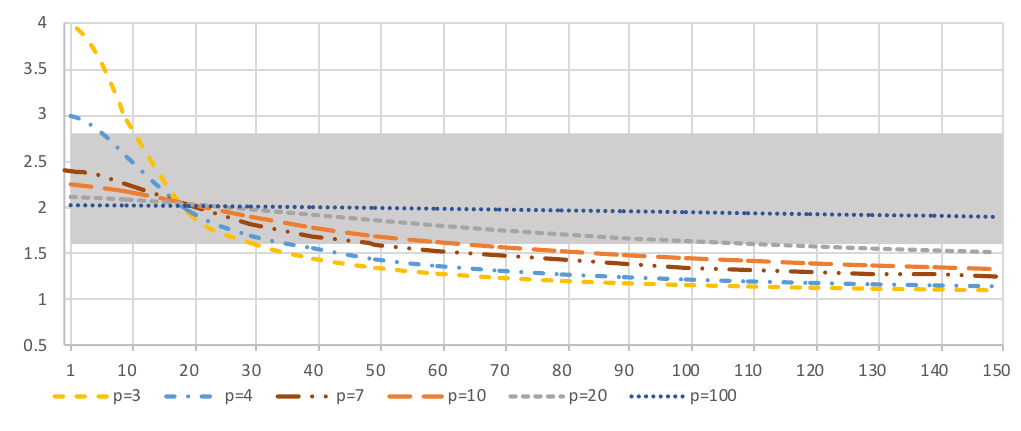}
\end{center}
\baselineskip=15pt
\begin{quote}
{\bf Fig. 14}~ The functional dependence of the scaling exponents $\b_r$ of the tensor ratio $r(N)$ on the energy scale $\mu$ at 
 $p=3,  4, 7, 10, 20, 100$ for hilltop (top panel) and hilltop-squared inflation (lower panel). The grey 
 region indicates the values for which the different models are compatible with the CMB and e-fold  based  constraints.
 \end{quote}
 
 \baselineskip=19.5pt

The results obtained here for hilltop and hilltop-squared inflation can now be compared with the results obtained 
in \cite{ls22} for modular inflation models.  For those models the $\mu$ range was restricted to the super-Planckian range 
$\mu > 10 M_\rmPl$ and the models showed a simple scaling-of-scaling behavior with exponents $\delta_r$ 
that were very similar, both close to $\delta_r \cong 2$. In order to compare the hilltop models to these modular 
inflation models it is useful to restrict them to the same energy range. This leads again to an approximate simple 
scaling behavior with exponents $\delta_r$ in dependence of $p$. These hilltop scaling exponents are much smaller 
than 2, establishing that the resulting scaling-of-scaling behavior can be used as a classification tool that distinguishes 
between these two different classes of models. The advantage of this analysis is that a single numerical exponent 
allows to characterize a model and its parameter space.

\vskip .3truein

\section{Conclusions} 

 In this paper we have shown how the energy dependent scaling exponents $\b_\cO(\mu)$ associated to 
observables $\cO(N)$ as functions of the number $N$ of inflationary e-folds  can be used to characterize the 
classes of hilltop and hilltop-squared inflation. Our focus has been on the 
 spectral tilt $\delta_n$ and the tensor-to-scalar ratio $r$,  leading to an analysis
 of the behavior of the scaling exponents $\b_{\delta_n}(\mu)$ and $\b_r(\mu)$ of
  $\delta_n(N)$ and $r(N)$. Here the energy scale $\mu$ is constrained by the conditions imposed, which 
  include the CMB and the e-fold constraint, but may also include other restrictions, for example the reheating 
  constraint. In the classes of hilltop and hilltop-squared inflation these constraints lead to $\mu$-ranges that increase 
  as the exponent $p$  is increased, leading to increasing domains of functions $\b_\cO$. 
  The resulting scaling exponents $\b_r(\mu)$ then lead to two different modes   in which models can be classified, 
  first via their characteristic intervals, and second via the functional types of the scaling 
  exponents. 
  
    The scaling results for the hilltop and hilltop-squared for $\b_\cO(\mu)$ are quite similar, as expected, indicating that 
  the differences seen by the inflaton toward the end of inflation do not lead to significant changes in the present context. 
  The behavior of these two classes is however quite different from that of the modular inflation models considered in \cite{ls22}. 
  Using the characteristic interval as a diagnostic tool shows that the modular inflation models are quite distinct from 
  hilltop inflation and hilltop-squared inflation since the tensor exponents of the former have no overlap with the 
  tensor exponents of the latter. Hence these models belong to different regions of the inflationary landscape.
   The functional behavior of hilltop and hilltop-squared models is also quite similar,
  but again quite different from that of the modular models. Comparing modular inflation with hilltop/hilltop-squared 
  inflation shows that the CMB + e-fold compatible range of $\mu$   for hilltop and hilltop-squared inflation is much 
  larger than for the modular inflation models. As discussed in section 9, for the full viable range the functional behavior of 
  the exponents depends quite dramatically on the model parameter $p$. 
  In order to compare HI and HSI models with modular inflation it is necessary to restrict the energy scale to 
  about $\mu \geq 10M_\rmPl$. In this range the functional behavior of HI and HSI simplifies and can be approximated 
  as a power law scaling relation, leading different  inflationary scaling regimes.
  
   An immediate implication of the energy dependence of the scaling exponents of both $\delta_n(N)$ and $r(N)$ 
 is that this  leads to an energy dependent scaling behavior for $r(\delta_n)$, leading to a family of curves in the 
  spectral-tensor plane. This will have consequences for the type of reheating analysis considered for example 
  in \cite{dm22}. 

The fact  that hilltop and hilltop-squared models lead to scaling regimes that are different from that of the modular inflation 
models considered in \cite{ls22} raises the question whether the scaling diagnostics considered here 
generalize to other models.  It would be interesting to place other inflationary models in the scaling framework 
considered here, either for singlefield inflationary models or for theories with more than one inflaton field. 
Analyzing models such as those considered in the singlefield encyclopedia \cite{mrv13}, or 
multifield models, including for example the models considered in the references
 \cite{l93, ht15, mm17, m19etal, abl19, apr19, c20etal, kt20,  bl22, l23, a23etal, k20, k23}, 
 will provide further insight into the scaling structure of the inflationary landscape.

\vskip .3truein

{\large {\bf Acknowledgement.} }\\
This work was supported in part by a sabbatical leave of absence. It is a pleasure to thank Ali Ishaq and Vipul Periwal for 
discussions and correspondence.

\end{document}